\documentclass{desyproc}

\begin{document}
%------------------------------------
\title{From Diphoton GDAs and Photon GPDs to the chiral odd Photon DA}

%for single authors the superscripts are optional
\author{{\slshape Lech Szymanowski$^{1,2}$}\\[1ex]
$^1$  So{\l}tan Institute for Nuclear Studies,
Ho\.za 69, 00-681 Warsaw, Poland \\
$^2$   LPT, Universit\'e d'Orsay, CNRS, 91404 Orsay, France  }

% if the proceedings are available online (e.g. at Indico)
% please enter the contribution ID or file_name below for the DOI
%\contribID{32}
\contribID{szymanowski\_lech}

% TO THE CONFERENCE EDITORS:
% please update the following information
% before sending the template to the authors
\confID{1407}  % if the conference is on Indico uncomment this line
\desyproc{DESY-PROC-2009-03}
\acronym{PHOTON09} % if you want the Acronym in the page footer uncomment this $
\doi  % if there is an online version we will register DOIs

\maketitle

\begin{abstract}
The photon is a very interesting object for QCD studies since it has both
a pointlike coupling to quarks, which yields a perturbative part of photonÕs wave function, and a 
non-perturbative coupling related to  the magnetic 
susceptibility of the QCD vacuum and which builds its
chiral-odd twist-2 distribution amplitude. The first feature allows us 
to compute
 the photon anomalous generalized parton distributions (GPD) and the 
diphoton generalized distribution amplitudes. The second feature allows us 
to use a transverse spin asymmetry to  probe the chiral odd 
distribution amplitude of the photon.
\end{abstract}

%\usepackage{verbatim}

%\usepackage{amsmath}
%\usepackage{amsfonts}
%\usepackage{amssymb}
%\usepackage{graphicx}
%\usepackage{bm}
%\usepackage{color}
%\usepackage{mathrsfs}\left[ \left\lbrace
%\usepackage{epsf}
%\usepackage{showlabels}
%\usepackage[hypertex]{hyperref}
%%%%%%%%%%%%%%%%%%%%%%%%
%\documentclass[prd,aps,twocolumn,superscriptaddress,epsf,amstex,amsmath,amssymb,draft,showpacs]{revtex4}
%%%%%%%%%%%%%%%%%%%%%%
%\documentstyle[preprint,aps,epsf,amstex]{revtex}
%\documentstyle[aps,twocolumn,epsf]{revtex}
%\documentstyle[aps,twocolumn,epsf,amstex]{revtex}
%\documentclass[aps]{revtex}
%\usepackage{aps,twocolumn,epsf,amstex}
%\documentstyle[aps,epsf]{revtex}
%\def\btt#1{{\tt$\backslash$#1}}
%\renewcommand{\baselinestretch}{1.0}
%\font\xxx=cmr12 scaled\magstep3
%\font\xxy=cmr12 scaled\magstep2
%\newcommand{\bin}[2]{C^{{#1}}_{{#2}}}

% \usepackage[english]{babel}
% \usepackage{pstricks}
%\usepackage{psfrag}
% \usepackage{epsfig,graphicx}
%\usepackage{epsf}
%\usepackage[dvips]{graphics}
% \usepackage{dcolumn}
% \usepackage{bm}

%%%%%%%%%%%%%% OUR NOTATION  %%%%%%%%%%%%%%%%%%

\newcommand{\Fq}{\Phi^{q,\, \Gamma}  }
\newcommand{\FA}{\Phi^{A,\, \Gamma}  }

%%%%%   couplings   %%%%%%%%%%%%%%%%%555555

\newcommand{\fV}{f_{3\,\rho}^V}
\newcommand{\fA}{f_{3\,\rho}^A}

\newcommand{\fr}{f_\rho}

\newcommand{\zV}{\zeta_{3}^V}
\newcommand{\zA}{\zeta_{3}^A}
%%%%%  polarization vectors %%%%%%%%%%

\newcommand{\rt}{e^{*\,T}}
\def\bea{\begin{eqnarray}}
\def\eea{\end{eqnarray}}
\def\beas{\begin{eqnarray*}}
\def\eeas{\end{eqnarray*}}
\def\beqas{\begin{eqnarray*}}
\def\eqas{\end{eqnarray*}}
\def\beq{\begin{equation}}
\def\eeq{\end{equation}}
\def\beqd{\begin{displaymath}}
\def\eeqd{\end{displaymath}}
\def\eqd{\end{displaymath}}

\def\slashchar#1{\setbox0=\hbox{$#1$}
   \dimen0=\wd0
   \setbox1=\hbox{/} \dimen1=\wd1
   \ifdim\dimen0>\dimen1
      \rlap{\hbox to \dimen0{\hfil/\hfil}}
      #1
   \else\begin{eqnarray}
      \rlap{\hbox to \dimen1{\hfil$#1$\hfil}}
      /
   \fi}

\section{Photon GPDs and diphoton GDAs}
The factorization of the amplitude for the deeply virtual Compton scattering (DVCS) process $\gamma^*(Q) \gamma \to \gamma \gamma$ at high $Q^2$ is demonstrated in two distinct kinematical domains, allowing to 
define the photon generalized parton distributions and the diphoton generalized distribution amplitudes. Both these quantities exhibit an anomalous scaling behaviour and obey new inhomogeneous QCD evolution equations.
The parton content of the photon has been the subject of many studies since the 
seminal paper by Witten \cite{Witten} which allowed to define the anomalous quark and gluon distribution
functions. Recent progresses in exclusive hard reactions  focus on generalized parton distributions (GPDs), which are defined as Fourier transforms 
of matrix elements between different states, such as $
\langle N'(p',s') |\bar\psi(-\lambda n)\gamma.n\psi(\lambda n) | N(p,s) \rangle$
and their crossed versions, the generalized distribution amplitudes (GDAs) which describe the exclusive hadronization of a $\bar q q $ or $g g$ pair 
in a pair of hadrons, see Fig.~1.
%\begin{center}
\begin{figure}
\hspace*{2.5cm}\includegraphics[width=9cm]{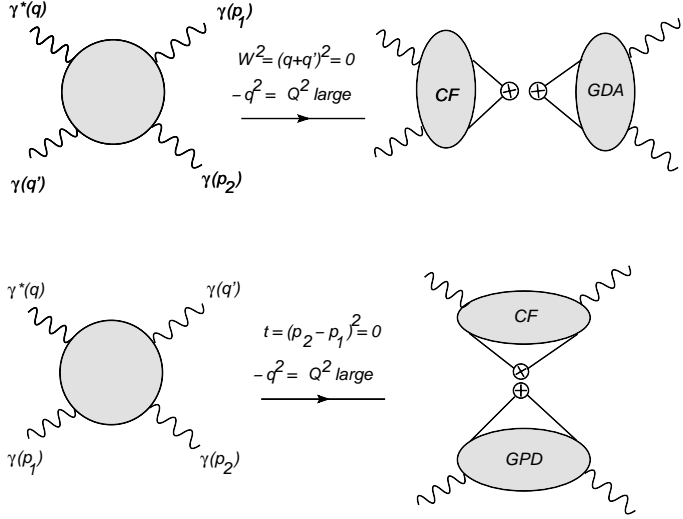}
\caption{Factorizations of the DVCS process on the photon.}
\end{figure}
%\end{center}
 In the photon case, these quantities are perturbatively calculable \cite{Friot,El} at leading order in $\alpha_{em}$ and leading logarithmic order in $Q^2$. They constitute an interesting theoretical laboratory for the non-perturbative hadronic objects that hadronic GPDs and GDAs are.

\subsection{The diphoton generalized distribution amplitudes}
Defining the momenta as $q= p - \frac{Q^2n}{s} ,~q'= \frac{Q^2n}{s} ,~p_1=\zeta p ,~p_2= (1-\zeta) p$, where 
$p$ and $n$ are two light-cone Sudakov vectors and $2 p\cdot n =s $,
the amplitude of the process
\begin{equation}
\gamma^*(Q, \epsilon) \gamma(q', \epsilon') \to \gamma(p_{1}, \epsilon_1) \gamma(p_2,\epsilon_2)
\label{dvcs}
\end{equation}
may be written as
$A = \epsilon_\mu\epsilon'_\nu{\epsilon_1}^*_\alpha{\epsilon^*_2}_\beta T^{\mu\nu\alpha\beta}$.
In forward kinematics where $(q+q')^2=0$,
the tensorial decomposition of $T^{\mu\nu\alpha\beta}$ reads (see \cite{El})
\begin{eqnarray}
&&
\frac{1}{4}g^{\mu\nu}_Tg^{\alpha\beta}_T W_1^q+
\frac{1}{8}\left(g^{\mu\alpha}_Tg^{\nu\beta}_T 
+g^{\nu\alpha}_Tg^{\mu\beta}_T -g^{\mu\nu}_Tg^{\alpha\beta}_T \right)W_2^q
+ \frac{1}{4}\left(g^{\mu\alpha}_Tg^{\nu\beta}_T - g^{\mu\beta}_Tg^{\alpha\nu}_T\right)W_3^q\, .
\end{eqnarray}
At leading order, the three scalar functions $W_i^q$ can be written in a factorized form which is particularly simple when the factorization scale $M_F$ equals the photon virtuality $Q$. $W_1^q$ is then the convolution 
$W^q_{1}= \int\limits_{0}^1 dz \, C_V^q(z) \, \Phi_1^q(z,\zeta,0)$ of the coefficient function
$ C_{V}^q = e_q^2\left(\frac{1}{z}-\frac{1}{1-z}\right) $
with the {\em anomalous} vector GDA ($\bar z = 1-z, \bar \zeta = 1 - \zeta$) :
\begin{eqnarray}
\label{Phi1}
&&\Phi_1^q(z,\zeta,0) =
 \frac{N_C\,e_{q}^2}{2\pi^2} \log{\frac{Q^2}{m^2}}
\left[\frac{\bar{z}(2z-\zeta)}{\bar{\zeta}}\theta(z-\zeta)+\frac{\bar{z}(2z-\bar{\zeta})}{\zeta}\theta(z-\bar{\zeta}) \right. \nonumber \\
&&\hspace*{4cm}+\left.\frac{z(2z-1-\zeta)}{\zeta}\theta(\zeta-z)+\frac{z(2z-1-\bar{\zeta})}{\bar{\zeta}}\theta(\bar{\zeta}-z)\right] .
\end{eqnarray}
Conversely, $W_3^q$ is the convolution  of the  function $C_{A}^q = e_q^2\left(\frac{1}{z}+\frac{1}{\bar z} \right)$
with the axial GDA :
\begin{eqnarray}
\label{Phi3}
\hspace*{-0.5cm}\Phi_3^q(z,\zeta,0) = \frac{N_C\,e_{q}^2}{2\pi^2} \log{\frac{Q^2}{m^2}}\left[\frac{\bar{z}\zeta}{\bar{\zeta}}\theta(z-\zeta)-\frac{\bar{z}\bar{\zeta}}{\zeta}\theta(z-\bar{\zeta})
- 
\frac{z\bar{\zeta}}{\zeta}\theta(\zeta-z)+\frac{z\zeta}{\bar{\zeta}}\theta(\bar{\zeta}-z)\right] \;
\end{eqnarray} 
and $W^q_2=0$.
Note that these GDAs are not continuous at the points $z =\pm \zeta$. The anomalous nature of $\Phi^q_1$ and $\Phi^q_3$ comes from their proportionality to $\log{\frac{Q^2}{m^2}}$, which reminds us of the anomalous photon structure functions. A consequence is that 
$\frac{d}{d \,\ln Q^2}\Phi^q_i \neq 0$; consequently the QCD evolution equations of the diphoton GDAs obtained with the help of the ERBL kernel are non-homogeneous ones.

\subsection{The photon generalized parton distributions}
We now look at the same process in different kinematics, namely  $q=-2\xi p+n ,~q'=(1+\xi) p ,~p_1=n ,~p_2= p_{1} + \Delta = (1-\xi) p$~,
where $W^2= \frac{1-\xi}{2\xi}Q^2$ and $t=0$.
The tensor
$T^{\mu\nu\alpha\beta}$ is now decomposed on different tensors  with the help of three  functions ${\cal W}^q_i$ as (see \cite{Friot}):
\begin{eqnarray}
\frac{1}{4}g^{\mu\alpha}_Tg^{\nu\beta}_T {\cal W}^q_1
+
\frac{1}{8}\left(g^{\mu\nu}_Tg^{\alpha\beta}_T 
+g^{\alpha\nu}_Tg^{\mu\beta}_T -g^{\mu\alpha}_Tg^{\nu\beta}_T \right){\cal W}^q_2
+ \frac{1}{4}\left(g^{\mu\nu}_Tg^{\alpha\beta}_T - g^{\mu\beta}_Tg^{\nu\alpha}_T\right){\cal W}^q_3\, 
\end{eqnarray}
 These functions can also be written in factorized forms which have direct parton model interpretations when  the factorization scale $M_F$ is equal to $Q$: 
${\cal W}_{1/3}^q=\int\limits_{-1}^1dxC_{V/A}^q(x)H_{1/3}^q(x,\xi,0)$, ${\cal W}_2=0$.
The coefficient functions are $C_{V/A}^q = - 2e_q^2\left(\frac{1}{x-\xi+i\eta} \pm \frac{1}{x+\xi-i\eta}\right)$ and the unpolarized $H_{1}^q$ and polarized $H_{3}^q$ anomalous GPDs of quarks inside a real photon read :
\begin{eqnarray}
\label{H1}
&&\hspace*{-1cm}H_{1}^q (x, \xi, 0) =
\frac{N_C\,e_{q}^2}{4\pi^2} \left[\theta(x-\xi)  \frac {x^2 + (1-x)^2-\xi^2}{1-\xi^2}\;\right. 
\nonumber\\
 &&\hspace*{1cm}
+ \left.\theta(\xi-x) \theta(\xi+x) \frac{x(1-\xi)}{\xi(1+\xi)} \;
- \theta(-x-\xi)  \frac{x^2+(1+x)^2-\xi^2}{1-\xi^2}\right]\; \ln\frac{Q^2}{m^2}\;,
\end{eqnarray}
\begin{eqnarray}
\label{H3}
&&\hspace*{-1cm}
H_{3}^q (x, \xi, 0) = \frac{N_C\,e_{q}^2}{4\pi^2} \left[\theta(x-\xi)   \frac {x^2 - (1-x)^2-\xi^2}{1-\xi^2}\;\right. \nonumber\\ &&\hspace*{1cm}
- \left.
\theta(\xi-x) \theta(\xi+x) \frac{1-\xi}{1+\xi} \;
+ \theta(-x-\xi)  \frac {x^2 - (1+x)^2-\xi^2}{1-\xi^2}\right]\; \ln\frac{Q^2}{m^2}\;.
\end{eqnarray}
Similarly as in the GDA case,  the anomalous generalized parton distributions $H_{i}^q$ are proportional to 
$\ln\frac{Q^2}{m^2}$, which violates the scaling. Consequently, the anomalous terms $H_{i}^q$ supply to the usual homogeneous DGLAP-ERBL evolution equations of GPDs a non-homogeneous term which changes  them into non-homogeneous evolution equations. 

We do not anticipate a rich phenomenology of these photon GPDs, but in the case of a high luminosity electron - photon collider which is not realistic in the near future. However, the fact that one gets explicit expressions for these GPDs may help to understand the meaning of general theorems such as the polynomiality and positivity \cite{pos} constrains or the analyticity structure \cite{ana}. For instance, one sees that a D-term in needed when expressing the photon GPDs in terms of a double distribution. One also finds that,  in the DGLAP region,  $H_1(x,\xi)$ is smaller than its positivity bound by a sizeable and slowly varying factor, which is of the order of $0.7 - 0.8$ for $\xi  \approx 0.3$.

\section{Accessing the photon chiral-odd DA and the proton transversity}
In Ref.\cite{PRL}, we describe a new way to access the photon distribution amplitude through the  photoproduction of lepton pairs on a transversally polarized proton.

The leading twist chiral-odd photon distribution 
amplitude $\phi_\gamma(u)$ reads \cite{Braun}
\begin{eqnarray}\label{def3:phi}
\langle 0 |\bar q(0) \sigma_{\alpha\beta} q(x) 
   | \gamma^{(\lambda)}(k)\rangle =  i \,e_q\, \chi\, \langle \bar q q \rangle
 \left( \epsilon^{(\lambda)}_\alpha k_\beta-  \epsilon^{(\lambda)}_\beta k_\alpha\right)  
 \int\limits_0^1 \!dz\, e^{-iz(kx)}\, \phi_\gamma(z)\,, 
\label{phigamma}
\end{eqnarray}    
where the normalization is chosen as $\int dz\,\phi_\gamma(z) =1$, 
and $z$ stands for the momentum fraction carried by the quark. The product of the quark condensate and of the magnetic susceptibility of the QCD vacuum
$\chi\, \langle \bar q q \rangle$ has been estimated \cite{BK} with the help of the QCD sum rules techniques to be of the order of 50 MeV  and a lattice estimate has recently been performed \cite{Bui}. The  distribution amplitude 
$\phi_\gamma(z)$ has a QCD evolution which drives it to an asymptotic form $\phi^{as}_\gamma(z) = 6 z (1-z)$.
Its $z-$dependence at non asymptotic scales is very model-dependent \cite{Bro}.

We consider the following process ($s_T$ is the transverse polarization vector of the nucleon):
\begin{equation}
\label{process}
\gamma(k,\epsilon) N (r,s_T)\to l^-(p)  l^+(p') X\,,
\end{equation}
with $q= p+p'$ in the kinematical region where $Q^2=q^2$ is large and the transverse component $ |\vec Q_\perp |$ 
of $q$ is of the same order as $Q$. 
Such a process  occurs either through a Bethe-Heitler amplitude (Fig. 2a) where the initial photon 
couples to a final lepton, or through Drell-Yan type amplitudes (Fig. 2b) where the final leptons originate from 
a virtual photon. 
Among these Drell-Yan processes, one must distinguish the cases 
where the real photon couples 
directly (through the QED coupling) to quarks or 
through its quark content.  
We thus consider the  contributions where the photon couples to the strong 
interacting particles through its lowest twist-2 chiral odd distribution 
amplitude (Fig. 2c and 2d). We will call this amplitude ${\cal A}_\phi$. 

One can easily see by inspection that interfering  the amplitude ${\cal A}_\phi$ with a pointlike amplitude is the only way to get at the level of twist 2 (and with vanishing quark masses) a contribution to  nucleon transverse spin dependent observables.
Reaction (\ref{process}) thus opens a natural access to the photon distribution amplitude \cite{BraunLech}, provided the amplitude 
${\cal A}_\phi$ interferes with the Bethe-Heitler or a  usual Drell-Yan process. 
Moreover, since this amplitude has an absorptive part,  {\em single spin} effects do not vanish. 
%%%%%%%%%%%%%%%%%%%%%%%%%%%%%%%%%%%%%%%%%%%%%%%%%%%%%%%%%%%%%%%%%%%%%%%%%
\begin{figure}
\includegraphics[width=3.2cm]{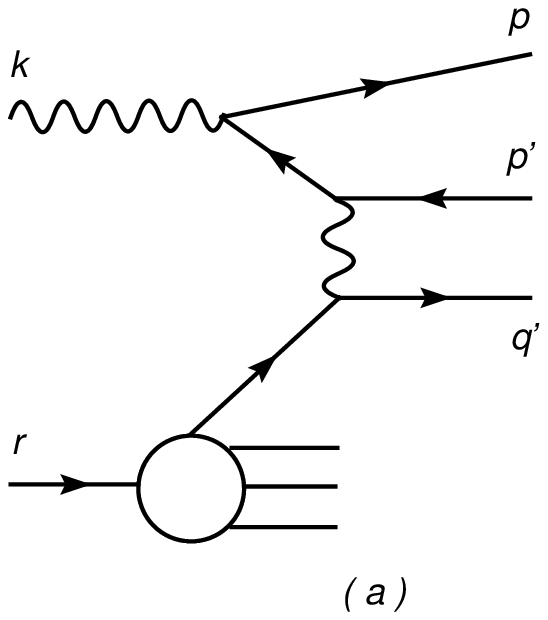}
\includegraphics[width=3.2cm]{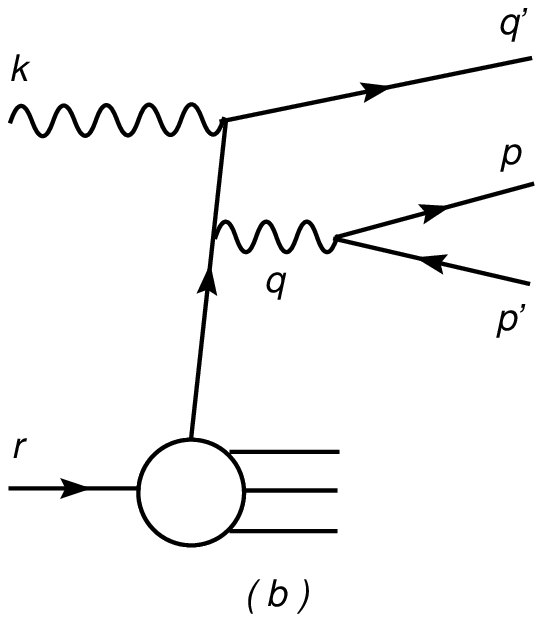}
\includegraphics[width=3.2cm]{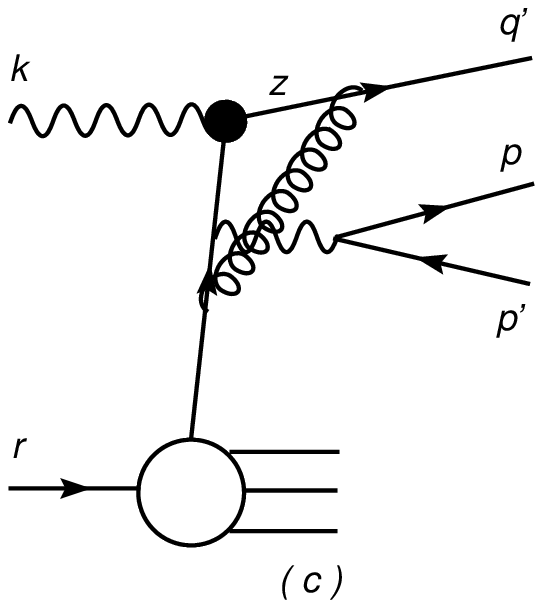}
\includegraphics[width=3.2cm]{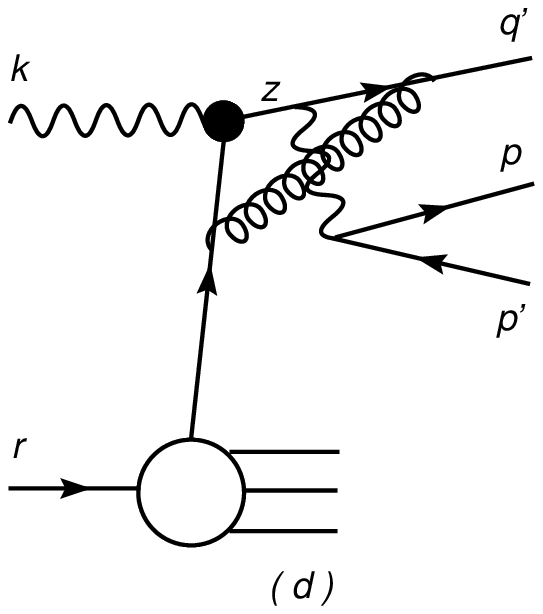}
\caption{Some amplitudes contributing to lepton pair photoproduction. (a) : The Bethe-Heitler process. (b) : The Drell-Yan process with the photon pointlike coupling. (c) -(d) : The Drell-Yan process with the photon Distribution Amplitude. }
\label{fig1}
\end{figure}
%%%%%%%%%%%%%%%%%%%%%%%%%%%%%%%%%%%%%%%%%%%%%%%%%%%%%%%%%%%%%%%%%%%%%%%%%
The amplitude where the photon interacts through its distribution amplitude  at lowest order (Fig. 2c and 2d) and in Feynman gauge, reads
\begin{eqnarray}\label{AmpC}
&&{\cal A}_\phi (\gamma q \to l \bar l q) 
\\
&&= 2i \frac{C_F}{4N_c} e_q^2 e 4\pi\alpha_s  \chi\, \langle \bar q q \rangle 
\frac{1}{Q^2} \int dz \phi_\gamma (z) 
 \bar u(q') [ \frac{A_1}{x\bar z s (t_1+i\epsilon)} +\frac{A_2}{ z u (t_2+i\epsilon)} ]u(r) \bar u(p)\gamma^\mu v(p') \,,
\nonumber
\end{eqnarray} 
with  $t_1= (zk-q)^2$ and $t_2= (\bar z k -q)^2$ and
\begin{eqnarray}\label{A12}
A_1 = x \,\hat r\,\hat \epsilon\,\hat k \,\gamma^\mu + \gamma^\mu \,\hat k \,\hat \epsilon \,\hat q \,,~~~~~~~~~~
A_2 = \hat \epsilon\,\hat q\, \gamma^\mu \,\hat k + \hat k \,\gamma^\mu \,\hat q \,\hat \epsilon \,,
\end{eqnarray}
which do not depend on the light-cone fraction $z$. Most interesting is the analytic structure of this amplitude since the quark propagators may be on shell
so that the amplitude ${\cal A}_\phi$ develops an absorptive part  proportional to
\begin{eqnarray}
\label{abs}
\int dz \phi_\gamma (z) \bar u(q') [\frac{A_1}{x\bar z s}\delta (t_1) +\frac{A_2}{ z u} \delta (t_2)]u(r) \bar u(p)\gamma^\mu v(p')\,. \nonumber
\end{eqnarray}
The $z-$integration, after using the $z-\bar z$ symmetry of the distribution amplitude, yields an absorptive part of the  amplitude ${\cal A}_\phi$ proportional to 
$\phi_\gamma (\frac{\alpha Q^2}{ Q^2+\vec Q_\perp ^2})$. This  absorptive part may be measured in single spin asymmetries and thus scans the photon chiral-odd distribution amplitude.

The cross section for reaction (\ref{process}) can  be decomposed as
\begin{eqnarray}\label{cs}
\frac {d\sigma}{d^4Q \,d\Omega}  =  \frac {d\sigma_{BH}}{d^4Q\,d\Omega} +  \frac {d\sigma_{DY}}{d^4Q\,d\Omega} +   \frac {d\sigma_{\phi}}{d^4Q\,d\Omega} +  \frac {\Sigma d\sigma_{int}}{d^4Q\,d\Omega}\,,\nonumber
\end{eqnarray}
where $\Sigma d\sigma_{int}$ contains various interferences, while
the transversity dependent  differential cross section (we denote $\Delta_T \sigma = \sigma(s_T) - \sigma(-s_T)$) reads
\begin{eqnarray}\label{cst}
&&\frac {d\Delta_T \sigma}{d^4Q\,d\Omega}  = \frac {d\sigma_{\phi int}}{d^4Q\,d\Omega}  \,,
\end{eqnarray}
 where $d\sigma_{\phi int}$ contains only interferences between the amplitude ${\cal A}_\phi$ and the other amplitudes. Moreover, one may use the distinct charge conjugation property (with respect to the lepton part) of the Bethe Heitler amplitude to select the interference between ${\cal A}_\phi$ and the Bethe-Heitler amplitude :
\begin{eqnarray}\label{CAcst}
&&\frac {d\Delta_T \sigma (l^-) - d\Delta_T \sigma (l^+) }{d^4Q\,d\Omega}  = \frac {d\sigma_{\phi BH}}{d^4Q\,d\Omega}  \,.
\end{eqnarray}
Conversely, one may use this charge asymmetry to cancel out the interference of ${\cal A}_\phi$ with the Bethe Heitler amplitude
\begin{eqnarray}\label{CScst}
&&\frac {d\Delta_T \sigma (l^-) + d\Delta_T \sigma (l^+) }{d^4Q\,d\Omega}  \propto \frac {d\sigma_{\phi DY}}{d^4Q\,d\Omega} \,.
\end{eqnarray}

The simplest  observable which contains all appealing features of our proposal  is the interference of   ${\cal A}_\phi$ and the Bethe-Heitler amplitudes, see Eq.\ref{CAcst}, in the unpolarized photon case. The polarization average of $d\sigma_{\phi BH}$ reads :
\begin{eqnarray}
\label{cspa}
&&\frac {1}{2} \sum_\lambda d\sigma_{\phi BH} (\gamma(\lambda)p\to l^-l^+X) 
\\
&&= \frac{(4\pi\alpha_{em})^3}{4 s } \, \frac{C_F 4 \pi\alpha_s}{2N_c}    
\frac{\chi\, \langle \bar q q \rangle}{\vec Q_\perp ^2}\int dx \sum_q Q_l^3Q_q^3h_1^q(x) 2{\cal R}e({\cal I}_{\phi BH})\,dLIPS \,,
\nonumber
\end{eqnarray}
with the usual phase space factor dLips
and
\begin{eqnarray}
\label{1}
{2\cal R}e({\cal I}_{\phi BH}) = \phi_\gamma [\frac{\alpha Q^2}{ Q^2+\vec Q_\perp ^2}]\, \frac{32\pi \alpha^2 \bar \alpha}{xs (\bar \alpha Q^2+\vec Q_\perp ^2)^2} { (Q^2 + \vec Q_\perp^2)} [\epsilon^{rks_TQ_T} {\cal{A}} + \epsilon^{rks_Tl_T} \cal {B}]\,,
\end{eqnarray}
where $\cal {A}$ and $\cal{B}$ are algebraic functions \cite{PRL}.
Eqs. \ref{cspa}, \ref{1} demonstate at the level of a highly differential cross section the existence of a non-vanishing observable proportional to  the photon distribution
 amplitude $\Phi_\gamma(z=\frac{\alpha Q^2}{ Q^2+\vec Q_\perp ^2})$ and the nucleon transversity $h_1$.

\paragraph*{Acknowledgements.}
\noindent
 This work is partly supported by the ECO-NET program, contract
18853PJ, the French-Polish scientific agreement Polonium and
the Polish Grant N202 249235.

\end{document}